# 10 Inventions on special type of keyboards
## -A study based on US patents

**Umakant Mishra**

Bangalore, India

umakant@trizsite.tk

http://umakant.trizsite.tk

**Contents**

# 1. Introduction

A keyboard is the most important input device for a computer. It is used with various types and sizes of computer. But the same standard keyboard will not work efficiently with different types of computers at different environments. There is a need to develop special keyboards to meet special requirements.

For example, the keyboard for young children can be different to facilitate easy learning. Comfy Interactive Movies Ltd developed a keyboard for young children which was known as ComfyKeyboard. The ComfyKeyboard has a set of large colorful picture keys. Similarly KidBoard, Inc. developed a KidBoard keyboard for children which had color-coded keys with pictures.

Similarly a keyboard in public place need to be more robust, the keyboard in a palmtop may have less number of keys, the keyboard for a game station may have special attachments, a multimedia keyboard may have CD ROM and speakers, a wireless keyboard may have remote control features, a touch sensitive keyboard may have sensors on the keys and so on.

## 2. Inventions on special type of keyboards

### 2.1 Coin operated personal computer with keyboard disable (5240098)

**Background**

A personal computer is useful for many purposes and it is desirable to provide computers at general public places (like, libraries, hotels, shopping centers, airports, schools etc.) for a nominal fee. One possibility is to use a coin operating mechanism like other coin-operated devices. But unlike other coin operated devices, the computer cannot be initialized as otherwise the user has to loose his unsaved work when the coin time is over.

**Solution provided by the invention**

The invention disables only the keyboard when the timer is over while other components of the computer system remain operational to retain the data. The user can insert another coin to get additional time to finish his work.

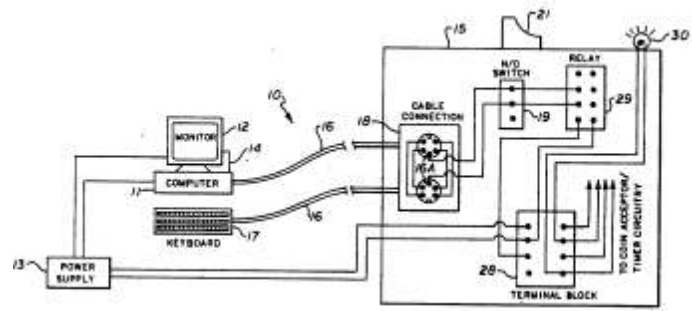

**TRIZ based analysis**

The objective is to provide personal computers at public places at nominal fee. The user should occupy if he has work and should not occupy when he does not pay **(Contradiction)**.

The invention provides personal computers at public places at cheap rates for a limited period of use **(Principle-27: Cheap and short living objects)**.

The conventional coin operated mechanism would initialize the computer machine to loose he unsaved work. Ideally the PC should stop working after the coin-period is over, but the user should not loose data **(Contradiction)**.

The invention just disables the keyboard while keeps the rest of the machine functional to preserve the work of the user **(Principle-20: Continuity of useful action)**.

**2.2 Hands-free hardware keyboard (5426450)**

**Background problem**

There are some limitations of manual cursor movements using the cursor keys on a keyboard. Moving the cursor keys requires a level of manual dexterity for their effective use. This is not suitable for physically disable people who don't have motor control organs. There is a need for a special device for the physically disabled users to use the computer without using the keyboard.

**Solution provided by the invention**

Donald Drumm invented a hands-free keyboard (patent 5426450, assigned to Wang Laboratories, Issued in June 95) which does not need any cursor keys to be pressed. The computer allows keyboard access in a hands-free environment. The user wears a headset with microphone and analog signal generators. The device has two operating modes, a cursor control mode and a keyboard simulation mode.

In the keyboard simulation mode, a microphone and preamplifier circuit is included together with the voice control system, which allows input selections to be made through voice commands. The user locates the desired keyboard input by observing the LED display and selects the input with voice commands.

In cursor control mode, the user controls the cursor movements by moving his head from leftward and rightward, forward and backward or different combinations of the both. The computer input device digitises the analog signals generated by the orientation.

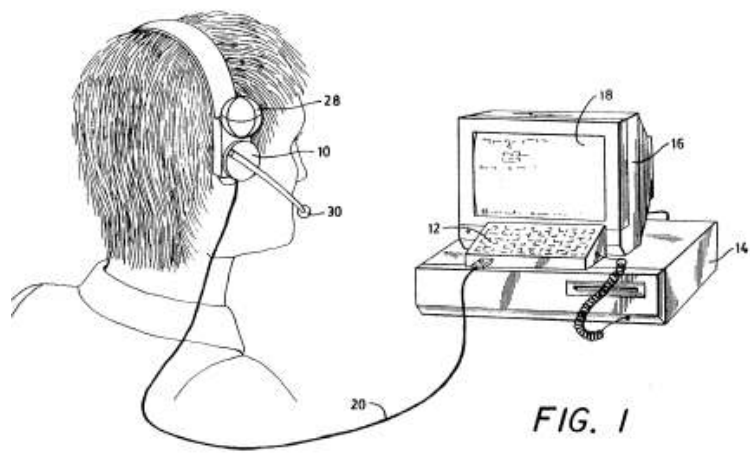

FIG. 1

**TRIZ based analysis**

The cursor should move by itself as per the wish of the user without the user physically moving a pointing device **(Ideal result)**.

The invention substitutes the mechanical action of pressing a key with a voice action **(Principle-28: Mechanics Substitution)**.

**2.3 Computer keyboard (5600313)**

**Background problem**

A keyboard is used as an input device for the computer. Although some keyboards contain function keys, their functions are different from software to software and hence difficult for the user to remember. There are no special keys on the keyboard to be used for specific jobs like printing, saving and other common tasks. There is a need for a keyboard, which contains predefined keys for various common functions.

**Solution provided by the invention**

Lorri Fredman invented (patent 5600313, issued Feb 1997) a keyboard that provides several mouse buttons on the keyboard. A set of static icon keys is positioned vertically on the left side of the keyboard. The command icon keys are positioned vertically on a right side of the keyboard. The tool bar icons are positioned horizontally above the function keys of the keyboard. These three sets of icon keys free up display screen which allows more screen space for the application.

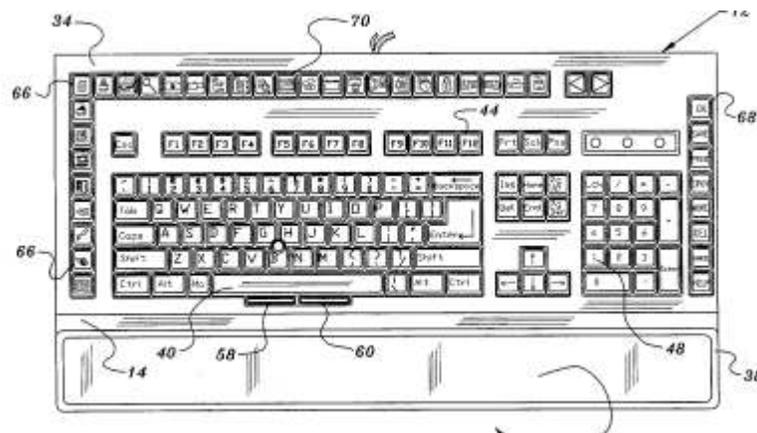

This keyboard provides all the advantages of a standard keyboard, besides it also provides the additional advantages of special icon keys.

**TRIZ based analysis**

The invention substitutes the visual toolbar icons with tactile keys on the keyboard **(Principle-28: Mechanics substitution)**.

The new keyboard provides additional keys over and above all the conventional keys **(Principle-38:Improve quality)**.

## 2.4 Keyboard controller with integrated RTC functionality (5854915)

### Background problem

The mobile chipsets were including a Real Time Clock (RTC) Functional System Block (FSB). RTC is used to update the time and date in the computer system. The RTC FSB has several problems, as it requires power and clock signal. It has limited programmability options. There is a need to provide an improved mobile chipset by removing RTC FSB from the mobile chipset and integrate RTC functionality into the keyboard controller.

### Solution provided by the invention

Goff et al. disclosed a keyboard controller with integrated Real Time Clock functionality (US patent 5854915, assigned to VLSI Technology, Issued in Dec 98). The keyboard controller with RTC functionality would be more efficient and make more efficient use of RTC CMOS RAM. The removal of RTC FSB from the mobile chipset would also reduce system cost.

### TRIZ based analysis

The invention moves the RTC FSB function and integrates into the keyboard controller **(Principle-5: Merging)**.

## 2.5 Wire/ wireless keyboard (5861822)

### Background problem

There are conventional wired keyboards and wireless keyboards as well. Both have their own limitations. The wired keyboard has limitation to work within a limited distance from the computer. The user cannot pull the keyboard beyond the length of the wire. A wireless keyboard overcomes this limitation but has other limitations. The wireless keyboards operate on batteries to transmit the data to the CPU. If the battery is down or not available then the keyboard is not usable.

### Solution provided by the invention

Samsung-Electro Mechanics invented a keyboard (US patent 5861822, Issued in Jan 1999) which can be used as a wired keyboard and a wireless keyboard depending on need. When the keyboard is connected to the computer, it will work as a conventional wire keyboard. When the cable is disconnected, the same keyboard works as a wireless keyboard. The keyboard can judge the connection status and automatically activates the desired mode. This facilitates the same keyboard to operate on dual mode and minimizes battery consumption.

### TRIZ based analysis

The same keyboard should work in both wire and wireless mode **(desired result)**. We need the keyboard to be wireless to take it far from the system. But we need the keyboard to be wired when we don't have battery in the keyboard. **(Contradiction)**.

The invention integrates the mechanism of both wired and wireless keyboards in the same module so that the same keyboard can work in both wired and wireless mode depending on the need **(Principle-40: Composite, Principle-6: Universality)**.

### 2.6 Transparent keyboard hot plug (5898861)

### Background problem

Conventionally while booting the PCs test the existence of a keyboard through a POST (power on self test) and if the keyboard is not present it gives a self-test failure and halts booting. This convention has a limitation, as sometimes it may be necessary to boot without a keyboard, as in case of a laptop using a stylus instead of a keyboard. Similarly a rack mounted server may not have a keyboard as it may have a keyboard only when necessary to do server level configuration.

### Solution provided by the invention

Emerson et al. found a solution to this problem (patent 5898861, assigned to Compaq Computer Corporation, April 99) by introducing a virtual keyboard. When the keyboard is plugged its presence is detected by the virtual keyboard controller. When the keyboard is unplugged, a virtual keyboard device communicates with the system keyboard controller (Principle- Cushioning and Principle- Mechanics Substitution). This provides a transparent plugging and unplugging of a keyboard independent of the system operation. When the keyboard is detected to be unplugged, the virtual keyboard is coupled with the system keyboard controller.

### TRIZ based analysis

The system should boot even without a keyboard as in some cases we need the system without a keyboard. But a keyboard is the primary input device; in most cases booting without a keyboard will lead to an unmanageable state of the system **(Contradiction)**.

The invention creates a virtual keyboard and makes the actual keyboard as an attachable and detachable device **(Principle-26: Copying, Principle-34: Discard and recover)**.

## 2.7 Redefining functions of a wireless remote Keyboard (5920308)

**Background problem**

A keyboard is used as an input device to the computer. In most implementations the data is transmitted from keyboard to computer. The computer does not send data to the keyboard. The same is true for the wireless keyboard. Typically the wireless keyboard contains the transmitter to transmit the signals and the main unit contains a receiver to receive the signals from the keyboard. This unidirectional communication does not allow to reconfigure keyboard or to reprogram a key-function.

**Solution provided by the invention**

Kwam-Wook Kim invented an improved keyboard (patent 5920308, assigned to Samsung Electronics, July 99) for PCs, which permits a bi-directional communication between the keyboard and the main computer through wireless remote control receivers. This allows the keyboard remote control receiver to receive information for redefining a fey function for remote control. As per the invention the key function for the remote control can be redefined in accordance with several sorts of programs.

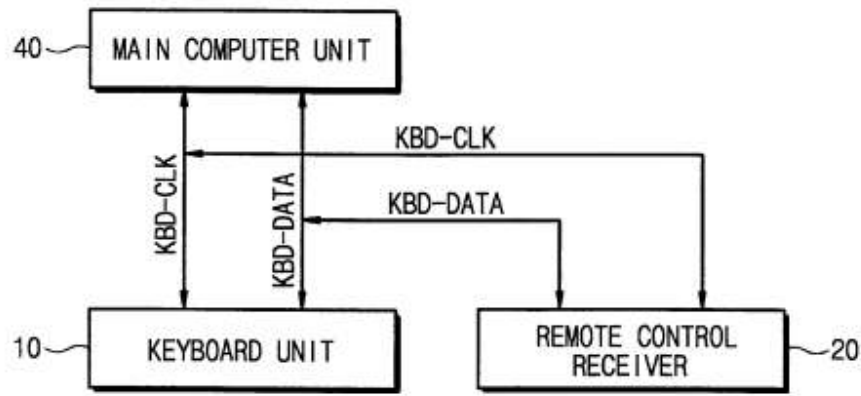

**TRIZ based analysis**

There is a need to improve the current keyboard mechanism to permit reprogramming a key-function. But there is no return communication path, especially in a remote control keyboard, from the computer to the keyboard. (Problem).

The present invention permits a bi-directional communication between the keyboard and computer so that the computer can send signals to the keyboard to re-program the keys **(Principle-17: Another Dimension, Principle-23: Feedback)**.

## 2.8 Computer keyboard systems and methods for determining excessive key stroke force (5982357)

### Background problem
People, who do extensive word processing, sometimes put excessive pressures on the keys that sometimes lead to Fatigue and Repetitive Stress Syndrome. There is a need to address this problem.

### Solution provided by the invention
Bugett et al. invented a keyboard system (patent 5982357, assigned to Key Tronic Corporation, Nov 99) that determines excessive keystroke force or overpressure.

According to the invention, the microprocessor is configured to determine a key switch closure duration value and compare that value with a predetermined threshold duration value. When the closure value is found to be higher than the predetermined value it generates a signal to warn the keyboard operator that excessive force has been applied. The operator will get cautious by the warning signal and will try to reduce his keystroke force.

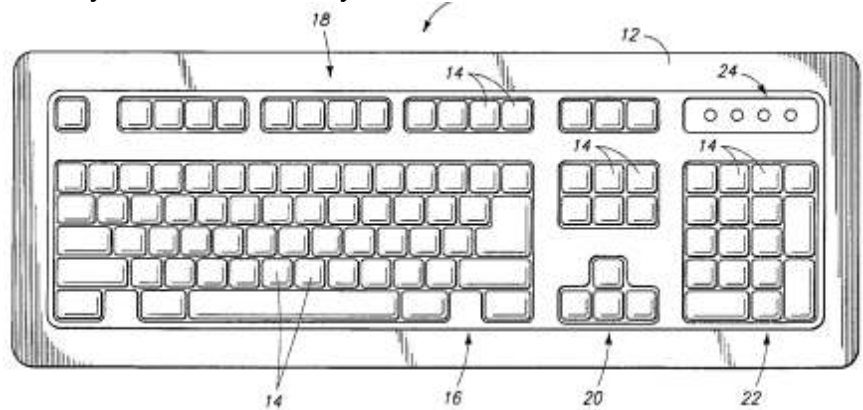

### TRIZ based analysis

The keyboard itself should discourage the operator to put overpressure **(Ideal Final Result)**.

The invention uses a method to calculate the force on the keys and uses a beep or warning message when the user puts overpressure **(Principle-23: Feedback)**.

## 2.9 Programmable multiple output force-sensing keyboard (5995026)

**Background problem**

Conventionally each key in a keyboard produces a specific signal, for example pressing 'a' produces a specific signal to display 'a' on the screen. Even the function keys also generate only one specific signal under operation. To change the signal of that key one has to combine the key with another special key like "ctrl-a" or "alt-a". This operation is confusing for the beginner and cumbersome for the experienced users. There is a need to selectively vary the output signals of a key on the keyboard.

**Solution provided by the invention**

Sellers invented a programmable keyboard sensitive to key depression force (US Patent 5995026, assigned to Compaq Computer Corporation, Nov 99). As per the invention the keyboard supports relatively different force applied by a user and measures the key depression force through a sensing resistor. The same key can generates different signals depending on the key depression force.

For example there can be a "low", "normal" and "high" force of depression which may generate three different types of signals. An example of its application may be, like a normal force will generate a small letter and a high force can generate a capital letter. The invention includes a method where a user can teach the keyboard to respond in a desired manner depending on the force of his keystroke.

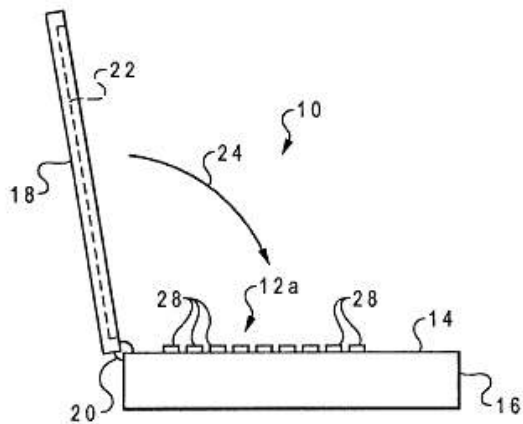

The invention also discloses a method to generate different signals depending on the duration of a depression. When the duration of depression is longer than the predefined period, the meaning of the key is altered. An example of the application of this method, if a high depression of a key generates a capital letter, a longer depression may cause it bold.

**TRIZ based analysis**

The invention uses the same key to generate different signals depending on the key depression force or key depression duration **(Principle-35: Parameter change)**.

### 2.10 Dynamic keyboard (6359572)

**Background problem**

The size of the notebooks and PDAs are getting smaller along with their keyboards. But a small keyboard is not convenient to operate because of the large size of human fingers. One of the solutions is to eliminate the full size physical keyboard and provide a soft keyboard on the touch sensitive display operated by finger or stylus. Even if the mechanism of the keyboard is changed, the size of the keyboard is still a limitation to display all the characters and symbols.

**Solution provided by the invention**

Peter Vale invented a programmable keyboard (Patent 6359572, assigned to Microsoft, Mar 02), which dynamically adjusts the meaning and appearance of one or more dynamic keys on a displayed keyboard in response to a user's predicted need. For example a single dynamic key may change to a different punctuation symbol in response to the predicted grammatical need.

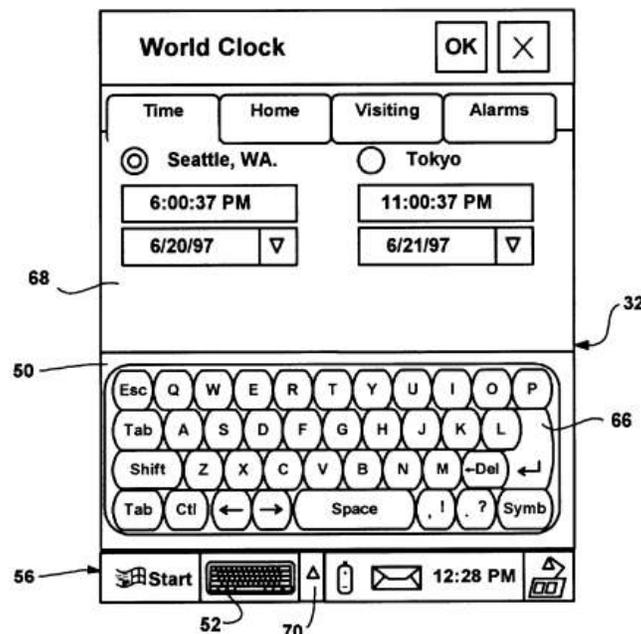

This dynamic nature of the keys eliminates the need of additional keys for different symbols and special characters for small size soft keyboards.

**TRIZ based analysis**

The same key should not be used for different symbols, as that would cause confusion with the user. On the other hand, the same key should change its meaning to appropriate alphabetic or punctuation symbols depending on the context so that there is no need for additional keys **(Contradiction)**.

The invention makes the soft keys programmable so that it can predict and display the relevant characters required by the user **(Principle-15: Dynamize)**.

The use of the same key for multiple symbols avoids the need for additional keys **(Princple-6: Universal)**.

## 3. Reference:

1. US Patent 5240098, "Coin operated personal computer with keyboard disable", inventors Desai et al., assignee- Nil, Issued Aug 1993.

2. US Patent 5426450, "Hands-free hardware keyboard", Donald Drumm, assigned to Wang Laboratories, June 95

3. US Patent 5600313, "Computer keyboard", invented by Lorri Freedman, issued Feb 1997

4. US Patent 5854915, Keyboard controller with integrated RTC functionality, Invented by Goff et al., assigned to VLSI Technology, Dec 98

5. US Patent 5861822, Keyboard working in both wire and wireless mode, Invented by Park et al., Assigned to Samsung Electro Mechanics, Jan 99

6. US Patent 5898861, "Transparent keyboard hot plug", Emerson et al., assigned to Compaq Computer Corporation, April 99

7. US Patent 5920308, "Redefining functions of a wireless remote keyboard", Kwam-Wook Kim, assigned to SamSung Electronics, July 99

8. US Patent 5982357, "Computer keyboard systems and methods for determining excessive key stroke force", Bugett et al., assigned to Key Tronic Corporation, Nov 99

9. US Patent 5995024, "Keyboard and notebook type computer", Kambayashi et al., assigned to Fujitsu Limited, Nov 99.

10. US Patent 6359572, "Dynamic keyboard"., Peter Vale, assigned to Microsoft, Mar 02

11. US Patent and Trademark Office (USPTO) site, http://www.uspto.gov/